\input{aipcheck}

\documentclass[
    ,final            % use final for the camera ready runs
  ]
  {aipproc}

\layoutstyle{6x9}

\begin{document}

\title{The Growth of the Stellar Seeds of Supermassive Black Holes}

\classification{97.60.Lf, 97.10.Bt, 97.10.Gz, 95.30.Ft, 95.30.Lz, 95.30.Jx}
\keywords      {black holes, supermassive stars, high redshift, galaxy formation}

\author{Jarrett L. Johnson}{
  address={Los Alamos National Laboratory, Los Alamos, NM  87545},
altaddress={Max-Planck-Institut f{\" u}r extraterrestrische Physik, 85748 Garching, Germany},
  email={jlj@lanl.gov}
}

\author{Bhaskar Agarwal}{
  address={Max-Planck-Institut f{\" u}r extraterrestrische Physik,
    85748 Garching, Germany}
}

\author{Daniel J. Whalen}{
  address={Los Alamos National Laboratory, Los Alamos, NM  87545}
  ,altaddress={Carnegie Mellon University, Pittsburgh, PA  15213} % additional visiting address
}

\author{Claudio Dalla Vecchia}{
  address={Max-Planck-Institut f{\" u}r extraterrestrische Physik,
    85748 Garching, Germany}
}

\author{Christopher L. Fryer}{
  address={Los Alamos National Laboratory, Los Alamos, NM  87545}
}

\author{Sadegh Khochfar}{
  address={Max-Planck-Institut f{\" u}r extraterrestrische Physik, 85748 Garching, Germany}
}

\author{Hui Li}{
  address={Los Alamos National Laboratory, Los Alamos, NM  87545}
}

\author{Mario Livio}{
  address={Space Telescope Science Institute, Baltimore, MD  21218}
}

\begin{abstract}
One of the most promising explanations for the origin of the billion
solar mass black holes (BHs) inferred to power quasars at redshifts
$z$ $\ge$ 6 is that supermassive stars (SMSs) with masses $\ge$ 10$^4$
M$_{\odot}$ collapse to form the seed BHs from which they grow.  Here we review recent
theoretical advances which provide support for this scenario.
Firstly, given sufficiently high accretion rates of gas into
the cores of primordial protogalaxies, it appears that neither the
high energy radiation emitted from the stellar surface nor the limited
lifetime of SMSs can prevent their growth to masses of up to $\ge$
10$^5$ M$_{\odot}$.  Secondly, recent cosmological simulations suggest
that the high fluxes of molecule-dissociating radiation which may be required in order to
achieve such high accretion rates may be more common in the early universe
than previously thought.  We conclude that the majority of
supermassive BHs may originate from SMSs at high redshifts.
\end{abstract}

\maketitle

%%%%%%%%%%%%%%%%%%%%%%%%%%%%%%%%%%%%%%%%%%%%
%% MAINMATTER
%%%%%%%%%%%%%%%%%%%%%%%%%%%%%%%%%%%%%%%%%%%%

\section{Introduction: The Rapid Growth of Black Holes}
There is growing observational evidence that the formation and growth
of black holes (BHs) in the earliest galaxies was, in many cases, very rapid.
Particularly strong constraints on the pace at which early BHs grow
comes from observations of quasars at $z$ $\ge$ 6 which are inferred to be
powered by accretion onto BHs with masses $\ge$ 10$^9$ M$_{\odot}$
(e.g. Mortlock et al. 2011).  Given the complexity of the formation
process of the early galaxies which harbored the first BHs, the
intense radiative feedback associated with their accretion, 
and the large amount of material that they must accrete in such a
short time, it is a challenge to understand the origin of these objects (e.g. Volonteri 2010).

One promising explanation for how BHs grow to supermassive scales by the time the
universe is only a few percent of its present age is that the 
BH seeds from which they grew were initially relatively massive themselves.
This is especially true if the rate at which BHs can grow is
limited by the radiation emitted in the accretion process (e.g. the
Eddington limit), as in this case smaller BHs cannot grow as 
quickly as larger ones.  As such seed BHs were likely born from the
collapse of primordial stars, this raises the question of how
massive such stars can become.  As we discuss below, there is growing
theoretical evidence for a population of supermassive
stars (SMSs), with masses $\ge$ 10$^4$, that are strong candidates for
the seeds of SMBHs.

\begin{figure}
  \includegraphics[height=.32\textheight]{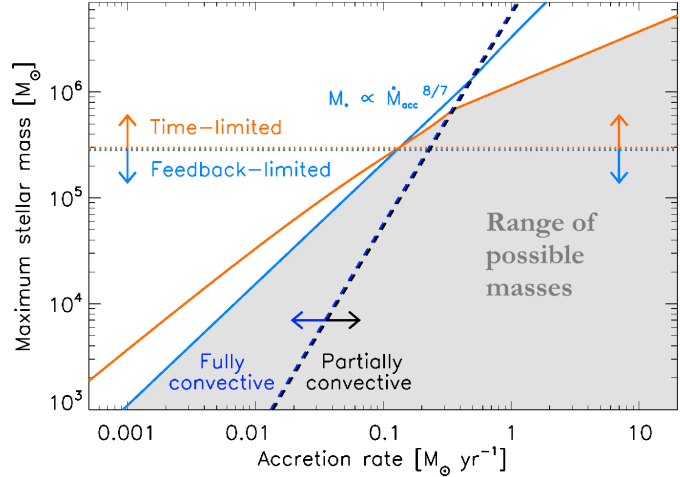}
  \caption{Maximum mass $M_{\rm *}$ to which a star can grow at a constant
    accretion rate $\dot{M}_{\rm acc}$. Above the maximum mass $M_{\rm
      *}$ $\propto$ $\dot{M}_{\rm acc}^{8/7}$ ({\it blue line}),
    radiative feedback shuts off accretion because the H~{\sc ii} region of
the star breaks out and prevents gas
infall. For accretion rates $\dot{M}_{\rm acc}$ $\ge$ 10$^{-1}$ M$_{\odot}$ yr$^{-1}$,
the maximum mass is set instead by the lifetime of the star ({\it red
line}). From Johnson et al. (2012a).}
\end{figure}

\section{The Formation of Supermassive Stars}
From simple theoretical arguments, the
characteristic mass of stars is thought to increase with the
temperature of the gas from which they form.
As the first, Population (Pop) III stars formed from gas cooled only
by the relatively inefficient coolant molecular hydrogen (H$_{\rm 2}$)
to $\simeq$ 200 K (an order of magnitude higher than that of
metal-enriched, star-forming gas today), they were likely relatively massive
(e.g. Bromm \& Larson 2004). Furthermore, in regions of the early universe where the
formation of H$_{\rm 2}$ was suppressed, for example by a large flux
of H$_{\rm 2}$-dissociating stellar radiation, the
gas could only cool via atomic hydrogen down to $\simeq$ 10$^4$ K
(e.g. Bromm \& Loeb 2003; Regan \& Haehnelt 2009).
Cosmological simulations following the collapse of such hot primordial
gas in protogalaxies suggest that the objects that form accrete at rates $\ge$
0.1 M$_{\odot}$ yr$^{-1}$ (Wise et al. 2008; Shang et al. 2010) and
may become SMSs with final masses of $\ge$ 10$^5$.  

This picture, however, is complicated by the strong ionizing radiation
that such SMSs would emit once they enter the main 
sequence; at some point, the increasing intensity of the radiative
feedback from the star should halt the flow of gas onto its surface
and limit its growth (e.g. Johnson et al. 2012a).  A second effect
which may limit the growth of SMSs is their short lifetime; once
their nuclear fuel is exhausted, they cannot grow any longer and 
will collapse to form the seed BHs discussed above (e.g. Begelman
2010).  As shown in Fig. 1, recent analytical calculations suggest that the first effect limits the growth of
SMSs which accrete at rates $\le$ 0.1 M$_{\odot}$ yr$^{-1}$ to $\sim$
10$^5$ M$_{\odot}$, but that the most rapidly accreting SMSs will
only be limited by the second effect and may grow to larger masses
(Johnson et al. 2012a).  Thus, given the extremely high accretion rates
rates found in cosmological simulations, it appears that little can
prevent the growth of SMSs up to $\sim$ 10$^5$ - 10$^6$ M$_{\odot}$, and their subsequent collapse to seed BHs 
of similar masses.

\begin{figure}
  \includegraphics[height=.225\textheight]{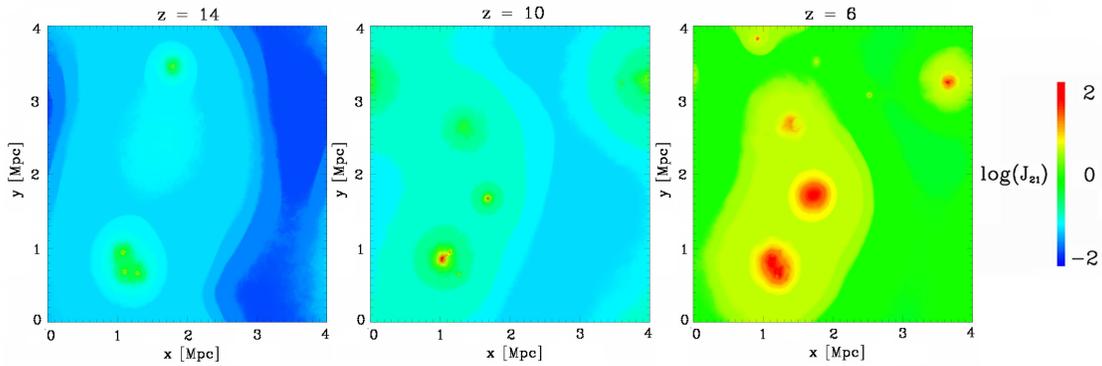}
  \caption{The H$_{\rm 2}$-dissociating flux  ($J_{\rm 21}$) generated by Pop~II and
    III stars in a simulated cosmological volume, at redshifts $z$ = 14 ({\it left}), 10 ({\it middle}), and 6 ({\it
  right}).  SMSs are expected to form in the cores of massive primordial halos that are
    exposed to fluxes of at least $J_{\rm 21}$ $\simeq$ 30 - 10$^3$.  From Johnson et al. (2012b).}
\end{figure}

\section{The Ubiquity of Supermassive Stellar Seeds}
As mentioned above, SMSs may only form in regions of the
early universe where an especially high flux of H$_{\rm
  2}$-dissociating radiation is produced, mostly likely by stellar sources.  Expressed in units
of 10$^{-21}$ erg s$^{-1}$ cm$^{-2}$ Hz$^{-1}$ sr$^{-1}$, the flux
required has been estimated to be in the range $J_{\rm 21}$ $\sim$ 30 - 10$^3$, with 
higher or lower values applying depending on the mass and metallicity
of the stars producing the flux (e.g. Shang et al. 2010).  While these
fluxes may be well above the average expected to pervade the 
universe at $z$ $\ge$ 6 (but see Petri et al. 2012), recent N-body
cosmological simulations suggest that in clustered regions they 
can be generated locally by early metal-enriched 
stellar clusters (Agarwal et al. 2012).  Indeed, these results suggest
that the conditions for SMS formation may be much more common in the
early universe than previously thought, perhaps leading to the
majority of SMBHs being seeded by SMSs.  As shown in
Fig. 2, recent cosmological hydrodynamics simulations accounting
explicitly for metal enrichment, Pop~II and III star
formation, and the locally- and globally-generated H$_{\rm 2}$-dissociating
radiation have corroborated this basic finding (Johnson et al. 2012b).

\begin{figure}
  \includegraphics[height=.3\textheight]{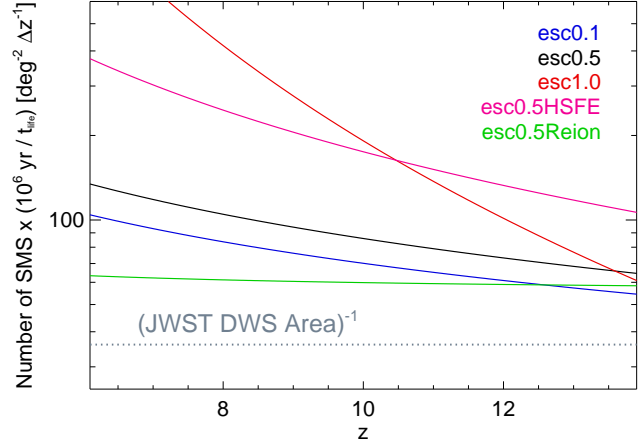}
  \caption{The number of supermassive stellar progenitors of direct
    collapse black holes, as observed on the sky per square degree
    per redshift interval ($\Delta$$z$), as a function of redshift $z$. The gray dotted
    line shows the number of SMSs that must be present
    for at least one per redshift interval ($\Delta$$z$ = 1) to appear in the
    field of view of the Deep-Wide Survey planned for the JWST. The
    survey should be large enough for up to
    of the order of 10 SMSs to lie within the field of
    view. From Agarwal et al. (2012).}
\end{figure}

\section{Summary and Outlook}
The two main conclusions of the work presented here are the following:

\begin{itemize}

\item Given the high accretion rates of gas into
  the cores of protogalaxies found from cosmological
  simulations, it appears that neither radiative
  feedback nor the limited lifetime of SMSs prevents
  their growth up to $\sim$ 10$^5$ - 10$^6$ M$_{\odot}$ (Johnson
  et al. 2012a).

\item The conditions under which such high accretion rates can be
  generated may be much more common than previously
  assumed, leading perhaps to the seeding of a majority of
  SMBHs by SMSs (Agarwal et al. 2012;
  Johnson et al. 2012b).

\end{itemize}

Ultimately, these findings will have to be verified by observations,
such as those scheduled to be conducted by the {\it James Webb Space
  Telecope} (JWST).  As Fig. 3 shows, Agarwal et
al. (2012) find that SMSs may form at sufficiently high rates for 
of the order of 10 to appear in the deep survey fields of the JWST.  Also, due to the high accretion
rates of primordial gas onto these objects, they are predicted to have
unique observable signatures.  Among these are
strong Balmer series and He~{\sc ii} $\lambda$1640 emission, as well
as comparatively weak Ly$\alpha$ emission (Johnson et al. 2012a).

%%%%%%%%%%%%%%%%%%%%%%%%%%%%%%%%%%%%%%%%%%%%%%%%
%% BACKMATTER
%%%%%%%%%%%%%%%%%%%%%%%%%%%%%%%%%%%%%%%%%%%%%%%%

\begin{theacknowledgments}
JLJ thanks the organizers of First Stars IV for the
opportunity to present this work, much of which was 
generously supported by the LANL LDRD program.
\end{theacknowledgments}




%\bibliography{sample}

%  \typeout{******************************************}
%  \typeout{** Please run "bibtex \jobname" to optain}
%  \typeout{** the bibliography and then re-run LaTeX}
%  \typeout{** twice to fix the references!}
%  \typeout{******************************************}

\end{document}